\documentclass[a4paper,11pt]{article} 
\usepackage{authblk}
\usepackage{fullpage}
\usepackage{graphicx}
\usepackage{xcolor}
\usepackage{algorithm}
\usepackage{algpseudocode}
\usepackage{amssymb}
\usepackage{amsthm}
\usepackage{amsmath}
\usepackage{mathrsfs}
\usepackage[mathcal]{euscript}
\usepackage{bm}
\usepackage{mathabx}
\usepackage[colorlinks,urlcolor=cobalt,citecolor=cobalt,linkcolor=cobalt,pdftex, pdfauthor = Lukas Exl]{hyperref}
\usepackage{tabularx}
\usepackage{setspace} 
\usepackage{wasysym}

\definecolor{airforce}{rgb}{0.16,0.32,0.75}
\definecolor{cobalt}{rgb}{0.0,0.28,0.67}

\newcommand*\luk[2]{{\color{black}{#2}}} 

\newtheorem{defi}{Definition}

\title{\Large{\textbf{Learning time-stepping by nonlinear dimensionality reduction to predict magnetization dynamics}}}
\author[1,4]{Lukas Exl \thanks{\texttt{lukas.exl@univie.ac.at}}}
\author[1,4]{Norbert~J.~Mauser}
\author[2,4]{Thomas Schrefl}
\author[3,4]{Dieter Suess}
\affil[1]{\small Wolfgang Pauli Institute c/o Faculty of Mathematics, University of Vienna, Austria.} 
\affil[2]{Department of Integrated Sensor Systems, Danube University Krems, Austria}
\affil[3]{Christian Doppler Laboratory of Advanced Magnetic Sensing and Materials, Faculty of Physics, University of Vienna, Austria }
\affil[4]{University of Vienna Research Platform MMM Mathematics - Magnetism - Materials, University of Vienna, Austria}

\begin{document}
%
\maketitle
\date

\noindent\textbf{Abstract.}  We  establish a time-stepping learning algorithm and apply it to predict the solution of the partial differential equation of motion in micromagnetism as a dynamical system depending on the external field as parameter. 
The data-driven approach is based on nonlinear model order reduction by use of kernel methods for unsupervised learning, yielding a predictor for the magnetization dynamics 
without any need for field evaluations after a data generation and training phase as precomputation. 
Magnetization states from simulated micromagnetic dynamics associated with different external fields are used as training data to learn a low-dimensional representation in so-called feature space 
and a map that predicts the time-evolution in reduced space. Remarkably, only two degrees of freedom in feature space were enough to describe the nonlinear dynamics of a thin-film element.
The approach has no restrictions on the spatial discretization and might be useful for fast determination of the response to an external field. 

\noindent\textbf{Keywords.} nonlinear model order reduction, kernel principal component analysis, kernelization, machine learning, micromagnetics\\

\noindent\textbf{Mathematics Subject Classification.} 	37M05, \, 62P35, \, 65Z05 

\section{Introduction}
Computational micromagnetism is nowadays used in many applications, like permanent magnets \cite{fischbacher2018micromagnetics} or magnetic sensors \cite{suess2018topologically}. 
A common numerical main task is to solve the Landau-Lifschitz-Gilbert (LLG) equation describing the motion of the magnetization in a magnetic material. Amongst other numerical efforts this involves many time-consuming 
computations of solutions to a Poisson equation in whole space \cite{abert2013numerical} for evaluating derivatives in the course of the time-stepping scheme \cite{miltat2007numerical,schrefl2007numerical}. 
On the other hand, electronic circuit design and real time process control need models that provide the sensor response quickly. Reduced order models in micromagnetism were mostly (multi)linear, e.g., 
tensor methods \cite{exl2014thesis} or based on spectral decomposition \cite{d2009spectral,bruckner2019large}. 
A linear notion of model order reduction was for instance introduced in \cite{d2009spectral}. The authors construct a subset of the eigenbasis of the discretized self-adjoint effective field operator. 
They show good compressibility of magnetic states of a dynamic thin-film benchmark switching problem  \cite{mumag4} when projected onto the linear subspace spanned by only a few eigenmodes. A main feature 
of this approach is the obtained simplification for the evaluation of the effective field resulting in a coupled system for the coefficients of the eigenmode expansion. 
However, in the original work the states where obtained from full micromagnetic simulations and latest numerical experiments show that reduced subspace integration turns out to need several hundreds of eigenmodes. 
This might still lead to a significant model reduction useful for large-scale applications. 
In addition, on-going research combines this linear model reduction with nonlinear projection such as Lambert projection indicating significant improvements concerning the number of needed eigenfunctions. The question of 
selecting the right subset of eigenfunctions is open and could be tackled with data-driven approaches, which is an on-going research line.\\ 
These recent advances give rise to advantages when using nonlinear approaches for model reduction. Very recently a procedure to \textit{learn the magnetization dynamics} for a range of different field values was introduced 
\cite{kovacs2019learning}. 
The conception is different for such methods: a rather large amount of data (magnetization states from dynamic simulations) is precomputed to train in a second step a predictor model on the basis of \textit{machine learning techniques}. 
The method in \cite{kovacs2019learning} trains auto-encoder and decoder by using \textit{convolutional neural networks (CNN)} yielding latent space approximations of magnetic states. In a second phase a 
feed-forward neural network is trained for 
predicting future states in latent space based on several previous states in a notion mimicking multi-step schemes for numerical integration of dynamical systems.\\ 
The inspiration of the present paper is to construct a time-stepping predictor on the basis of a non-black-box nonlinear dimensionality reduction approach for the better understanding of the underlying approximations. We will ground our approach on the mathematical theory of \textit{kernel methods} (\textit{kernelization}) that is well known 
for its success as \textit{nonlinear dimensionality reduction} in \textit{unsupervised learning}. A time-stepping predictor is constructed by learning a map between higher dimensional spaces called the \textit{feature spaces}, where 
the magnetic states can be well approximated by using only few components of a \textit{nonlinear principal component analysis}. The overall trained estimator is capable to predict the nonlinear magnetization dynamics of 
the benchmark \cite{mumag4} with only two degrees of freedom in feature space.\\   
The paper is organized as follows. In the next section we introduce kernel methods and feature spaces, specifically the nonlinear version of the  principal component analysis will be derived as well as the 
methodology of learning maps via the use of kernels. In section~\ref{sec:timestep} we will apply the approach to establish a learning approach for time-stepping of the micromagnetic parameter-dependent dynamical system. 
In section~\ref{sec:numerics} the scheme is numerically validated for the micromagnetic benchmark problem no.4 \cite{mumag4}.

\section{Nonlinear models in feature space}
Suppose $m\in\mathbb{N}$ given data points $x_k \in \mathbb{R}^N,\,k=1,\hdots,m$ . The \textit{(centered) covariance matrix} of the data is defined as 
$C := \frac{1}{m} \sum_{j=1}^m(x_j - \langle x \rangle)(x_j - \langle x \rangle)^T$, where $\langle x\rangle$ denotes the mean value. Calculating its eigenvectors, the \textit{principal axes}, results in an orthogonal system, where the 
corresponding eigenvalues equal the amount of variance in the respective principal direction. Coordinates in the eigensystem are denoted as \textit{principal components}. The system of eigenvectors associated with the largest $r$ eigenvalues encompasses the maximal possible amount of variance under all orthogonal systems of dimension $r$. 
The orthogonal basis transformation which maps a vector to its principal components is known as \textit{Principal Component Analysis (PCA)}, where we, in viewpoint of the forthcoming nonlinear extension, denote it as \textit{linear PCA} . PCA can be 
used for \textit{data compression} where only the largest principal components (corr.  to largest variance) are kept to conserve the most information. This notion is also a key in \textit{unsupervised learning}, e.g. \textit{manifold learning}, which aims at detecting relevant structure in data .\\
Linear PCA can not always reveal all relevant information and structure in the data. Therefore a \textit{nonlinear extension} has been introduced where the input data are first (possibly) nonlinearly mapped 
to a \textit{feature space} \cite{scholkopf1997kernel}.    
This is done via \textit{kernels}. 
\subsection{Principal component analysis in feature space}
We start with a brief introduction to \textit{kernels} and \textit{reproducing kernel Hilbert spaces (RKHS)}, for a more comprehensive discussion the interested reader 
is referred to a review of kernel methods in machine learning \cite{hofmann2008kernel}. 
\begin{defi}[Positive definite kernel function]\label{def:kernel} 
 Let $\mathcal{X}$ be a nonempty set. A symmetric function $k:\, \mathcal{X} \times \mathcal{X} \rightarrow \mathbb{R}$ is a positive definite kernel on $\mathcal{X}$ if 
 for all $m\in \mathbb{N}$ any choice of inputs $\mathbf{x} = \{x_1,\hdots,x_m\} \subseteq \mathcal{X}$ gives rise to a positive definite gram matrix $K[\mathbf{x}] \in \mathbb{R}^{m \times m}$ 
 defined as $K_{ij} = k(x_i,x_j),\,i,j = 1,\hdots,m$. 
\end{defi}
We will refer to positive definite kernels as \textit{kernels}. 

An important class of kernels are the \textit{Gaussian kernels}, here also referred to as \textit{radial basis functions (RBF)}.
\begin{defi}[RBF]\label{def:rbf}
 Let $\mathcal{X}$ be a dot product space. The radial basis function (RBF) kernel between two vectors $x,y \in \mathcal{X}$ is defined as 
 \begin{align}
  k(x,y) = e^{-\gamma \|x-y\|^2}.
 \end{align}
For the choice $\gamma = 1/\sigma^2$ the kernel $k$ is also known as the Gaussian kernel of variance $\sigma^2$.
\end{defi}
Kernels can be regarded as \textit{similarity measures} and thus they can be used to generalize structural analysis for data like the PCA.
Indeed, one can construct a (higher dimensional) \textit{Hilbert space} $\mathcal{F}_k$, which we call the \textit{feature space} of $\mathcal{X}$ associated with the kernel $k$, where the inner product is defined by the kernel $k$. 
This corresponds to mapping the data with some (possibly nonlinear) map $\Phi:\,\mathcal{X} \rightarrow \mathcal{F}_k$ such that the inner product in $\mathcal{F}_k$ of two mapped data points is given as 
\begin{align}\label{eqn:kerneltrick}
 \Phi(x) \cdot \Phi(y) = k(x,y).
\end{align}
Thus in feature space $\Phi(\mathcal{X}) = \mathcal{F}_k$ the similarity measure is "linearized". $\mathcal{F}_k$ is a reproducing kernel Hilbert space and mathematically well understood, 
see e.g. \cite{saitoh1988theory} for theoretical background. 
An intuitive way of thinking about the \textit{feature map} $\Phi$ is to imagine a mapped data point as a new vector $\Phi(x) = (p_1(x),p_2(x),\hdots)^T$ where the nonlinear functions $p_j$ define 
the coordinates of $\Phi(x)$ in the higher dimensional feature space. One can now try to learn structure via the mapped inputs. Linear algorithms for unsupervised learning, like the PCA, can be adapted to operate in feature space without the 
explicit knowledge of the underlying map $\Phi$, owing to the relation \eqref{eqn:kerneltrick}, also known as the \textit{kernel trick} in the machine learning community. The extension of linear PCA to its nonlinear 
analogue is known as \textit{kernel principal component analysis (kPCA)} \cite{scholkopf1997kernel}, which is the \textit{kernelized} version of linear PCA and given as follows.
\begin{defi}[kPCA]\label{def:kpca}
 Given inputs $\mathbf{x} = \{x_1,\hdots,x_m\} \subseteq \mathcal{X}$ and a kernel $k:\, \mathcal{X} \times \mathcal{X} \rightarrow \mathbb{R}$ the kernel PCA generates 
 kernel principal axes $v^{(j)} = \tfrac{1}{\sqrt{\lambda_j}} \,\sum_{i=1}^m \alpha_i^{(j)} \Phi(x_i),\, j=1,2,\hdots$ where the coefficient vectors $\alpha^{(j)} \in \mathbb{R}^{m}$ result from the eigenvalue problem 
 \begin{align}\label{eqn:eigkpca}
  G\alpha^{(j)} = m \lambda_j \alpha^{(j)},
 \end{align}
where the centered gram matrix $G = K[\mathbf{x}] - \mathbf{1}_m K[\mathbf{x}] - K[\mathbf{x}]\mathbf{1}_m + \mathbf{1}_mK[\mathbf{x}]\mathbf{1}_m \in \mathbb{R}^{m \times m}$ with $(\mathbf{1}_m)_{ij} = 1/m$ is used. 
The eigenvalue problem \eqref{eqn:eigkpca} is solved for nonzero eigenvalues. 
The $j$-th kernel principal component of a data point $x \in \mathcal{X}$ can be extracted by the projection  
\begin{align}
 p_j(x) = \Phi(x) \cdot v^{(j)} = \frac{1}{\sqrt{\lambda_j}} \sum_{i=1}^m \alpha_i^{(j)} k(x_i,x).
\end{align}
\end{defi}
For the purpose of \textit{nonlinear dimensionality reduction} only a few kernel principal components are extracted.

\subsection{Learning maps via kernels}\label{sec:learningmaps}
\begin{figure}[hbtp]
\centering 
\includegraphics[scale=1.5]{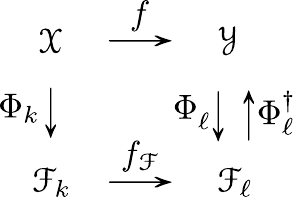}    
\caption{Illustration of the mappings in \eqref{eqn:mappings}.}\label{fig:mappings}
\end{figure}

A general \textit{learning problem} is to estimate a map between an input $x \in \mathcal{X}$ and output $y \in \mathcal{Y}$ from a given training set 
$(x_1,y_1), (x_2,y_2), \hdots, (x_m,y_m) \in \mathcal{X} \times \mathcal{Y}$. We denote $\mathcal{X}$ as the \textit{input set} and $\mathcal{Y}$ as the \textit{output set}. 
Mathematically, this refers to estimating the map $f$ from a \textit{hypothesis class} $\mathcal{H} = \{f(.;\alpha):\, \alpha \,\,\textrm{feasible parameter}\}$ by minimizing the \textit{risk function}, that is, 
\begin{align}
 r(\alpha) = \int_{\mathcal{X}\times \mathcal{Y}} L(y,f(x;\alpha))\, \textrm{d}\rho(x,y),
\end{align}
where $\rho$ is the unknown joint probability measure and $L:\,\mathcal{Y} \times \mathcal{Y} \rightarrow \mathbb{R}$ a \textit{loss function}. We can define $L$ as the distance in output feature space using 
a kernel $\ell:\,\mathcal{Y} \times \mathcal{Y} \rightarrow \mathbb{R}$ on the output set. 
Hence, there is a RKHS $\mathcal{F}_\ell$ with associated map $\Phi_\ell:\, \mathcal{Y} \rightarrow \mathcal{F}_\ell$ and $\ell(y,y^\prime) = \Phi_\ell(y)\cdot \Phi_\ell(y^\prime)$. 
Estimating a map $f:\,\mathcal{X} \rightarrow \mathcal{Y}$ according to the dependency of the available training data will require to minimize loss expressions $L(y,f(x)) = \|\Phi_\ell(y) - \Phi_\ell(f(x))\|_{\mathcal{F}_\ell}^2$, 
which can be expressed entirely through the kernel $\ell$ using the kernel trick. We note here that $\ell$ defined as the usual inner product $y^Ty^\prime$ (in the case $\mathcal{Y}$ is some subspace of an Euclidean space) 
would give $\Phi_\ell = \textrm{id}$, the identity. In the forthcoming numerics we will rather use a nonlinear kernel such as RBF to extract nonlinear features in output space.\\ 
The problem of estimating the map $f$ can be decomposed in subtasks using the idea of \textit{kernel dependency estimation (KDE)} \cite{weston2003kernel}. The map $f$ can be understood as the composition of three maps, that is,
\begin{align}\label{eqn:mappings}
 f = \Phi_\ell^{\dagger} \circ f_{\mathcal{F}} \circ \Phi_k, 
\end{align}
where $\Phi_k:\,\mathcal{X} \rightarrow \mathcal{F}_k$ is the feature map for inputs associated with a kernel $k$, $f_{\mathcal{F}}:\,\mathcal{F}_k \rightarrow \mathcal{F}_\ell$ the map between input and output feature spaces and 
$\Phi_\ell^\dagger:\,\mathcal{F}_\ell \rightarrow \mathcal{Y}$ an approximate inverse onto $\mathcal{Y}$ which is called the \textit{pre-image map}. 
See Fig.~\ref{fig:mappings} for an illustration of the involved mappings. For given inputs KDE learns first the map between feature spaces. 
This is done by estimating a (ridge) regression model from inputs to kernel principal components in feature space. 
More precisely, we determine the $r$ components of the kPCA (see Def.~\ref{def:kpca}) with kernel $\ell$ for each of the training data points $y_i,\,i=1,\hdots,m$, which results in the matrix $P \in \mathbb{R}^{m \times r}$ with entries 
$P_{ij} = p_j(y_i)$. Then we minimize the regularized linear model for $j=1,\hdots,r$
\begin{align}\label{eqn:ridgereg}
 \min_{b^{(j)} \in \mathbb{R}^{m}} \|P^{(j)} - K[\mathbf{x}] b^{(j)} \|^2_2+ \mu\,\|b^{(j)}\|_2^2,
\end{align}
where the superscript $(j)$ denotes $j$-th column and $K[\mathbf{x}]$ the gram matrix associated with kernel $k$ (input space).
We used the regularization parameter $\mu=0.1$ throughout the numerical experiments in section~\ref{sec:numerics}. 
The estimator of the image of the map $f_{\mathcal{F}}\circ\Phi_k $ for a new input $x\in\mathcal{X}$ is then given as 
\begin{align}
 \widehat{(f_{\mathcal{F}} \circ \Phi_k)}(x) = \big(\sum_{i=1}^m b^\ast_{i1} k(x_i,x),\hdots, \sum_{i=1}^m b^\ast_{ir} k(x_i,x)\big),
\end{align}
where $b^\ast \in \mathbb{R}^{m \times r}$ contains as columns the $r$ solutions to the problems \eqref{eqn:ridgereg}.\\ 
The output $y\in\mathcal{Y}$ of the new input pattern $x$ is then estimated from the learned image $\hat{f}_\mathcal{F}(x)$ 
and determined as approximate pre-image, that is, an approximation of a solution to the minimization problem
\begin{align}
 \min_{y \in \mathcal{Y}} \| \big(\Phi_\ell(y) \cdot v^{(1)},\hdots,\Phi_\ell(y) \cdot v^{(r)} \big) - \widehat{(f_\mathcal{F} \circ \Phi_k)}(x)\|^2.
\end{align}
This can be established using a supervised learning approach during training the kPCA in feature space $\mathcal{F}_\ell$ \cite{bakir2004learning}. 

\subsection{Learning time stepping for the Landau-Lifschitz-Gilbert equation}\label{sec:timestep}
The fundamental micromagnetic equation for dynamics in a magnetic body $\Omega \subset \mathbb{R}^3$ is the \textit{Landau-Lifschitz-Gilbert} (LLG) equation \cite{kronmueller}. 
Considering the magnetization as a vector field $\boldsymbol{M}(x,t) = M_s \boldsymbol{m}(x,t),\,|\boldsymbol{m}(x,t)| = 1$ 
depending on the position $x \in \Omega$ and time $t\in \mathbb{R}$ the LLG equation is given in explicit form as  
\begin{align}\label{eqn:LLGM}
 \frac{\partial \boldsymbol{M}}{\partial t} &\, = -\frac{\gamma_0}{1+\alpha^2} \, \boldsymbol{M} \times \boldsymbol{H} - \frac{\alpha\,\gamma_0}{(1+\alpha^2) M_s} \, \boldsymbol{M} \times \big(\boldsymbol{M} \times \boldsymbol{H}\big),
\end{align}
where $\gamma_0$ is the gyromagnetic ratio, $\alpha$ the damping constant and $\boldsymbol{H}$ the effective field, which is the sum of nonlocal and local fields such as the stray field 
and the exchange field, respectively, and 
the external field $\boldsymbol{h}\in\mathbb{R}^3$ with length $h$. The stray field arises from Maxwell's magnetostatic equations, that is, $\nabla\cdot \mathbf{H}_d = -\nabla\cdot \mathbf{M}$ in $\mathbb{R}^3$,  
whereas the exchange term is a continuous micro-model of Heisenberg exchange, that results in $\mathbf{H}_{ex} = \tfrac{2\,A}{\mu_0 M_s^2} \Delta \mathbf{M}$, where $\mu_0$ is the vacuum permeability, $M_s$ the saturation magnetization and $A$ the exchange constant.
Equation \eqref{eqn:LLGM} is a partial differential equation supplemented with an initial condition $\boldsymbol{M}(x,t=0) = \boldsymbol{M}_0$ and (free) Neumann boundary conditions. For further details of micromagnetism the interested reader is referred to the literature \cite{brown1963micromagnetics,aharoni2000introduction,kronmueller}.
Typically, equation \eqref{eqn:LLGM} is numerically treated by a semi-discrete approach \cite{suess2002time,donahue1999oommf,d2005geometrical,exl2017extrapolated}, 
where spatial discretization by collocation leads to a (rather large) system of ordinary differential equations. Clearly, the evaluation of the right hand side of the system is very expensive mostly due to the stray field, 
hence, effective methods are of high interest. The following \textit{data-driven} approach yields a predictor for the magnetization dynamics without any need for field evaluations after a data generation and training phase as precomputation. \\ 
For specific choice of the external field $h$ we consider the discretized unit magnetization components at time $t$ given as a vectors of length $N$, that is $m^{(p)}_h(t) \in \mathbb{R}^N$, $p=1,2,3$, where we assume 
$N$ degrees of freedom related to the spatial discretization. We consider a choice of field values $\{h_1,h_2,\hdots,h_m\} \subseteq H$, where the interval $H$ denotes some range of field values. 
Let us denote the set of magnetization components at time $t$ corresponding to the choice of field values to be denoted as $\mathcal{M}^{(p)}_{\{h_1,\hdots,h_m\}}(t) = \{m^{(p)}_{h_1}(t),m^{(p)}_{h_2}(t),\hdots, m^{(p)}_{h_m}(t)\} \subseteq \mathcal{M}_{H;t}^{(p)}$, 
where $\mathcal{M}_{H;t}^{(p)} \subseteq \mathbb{R}^N$ is the set of magnetization components at $t$ resulting from the initial value problem \eqref{eqn:LLGM} for $h \in H$. 
Assuming the next time point at $t + \Delta t,\, \Delta t > 0$, our task is to learn maps $f_{t,t + \Delta t}^{(p)}:\,\mathcal{M}_{H;t}^{(p)} \rightarrow \mathcal{M}_{H;t+\Delta t}^{(p)},\, p=1,2,3$ from the available data $\mathcal{M}_{\{h_1,\hdots,h_m\}}^{(p)}(t)$ and $\mathcal{M}_{\{h_1,\hdots,h_m\}}^{(p)}(t+\Delta t)$.   
For this purpose we follow the approach of the previous section~\ref{sec:learningmaps}, where we additionally append the respective field parameters (with some scaling) to each element in the data set $\mathcal{M}^{(p)}_{\{h_1,\hdots,h_m\}}(t)$. We next apply the procedure to the benchmark problem \cite{mumag4}. 

\section{Application to micromagnetism}\label{sec:numerics}
\begin{figure}[hbtp]
\centering 
\includegraphics[scale=0.45]{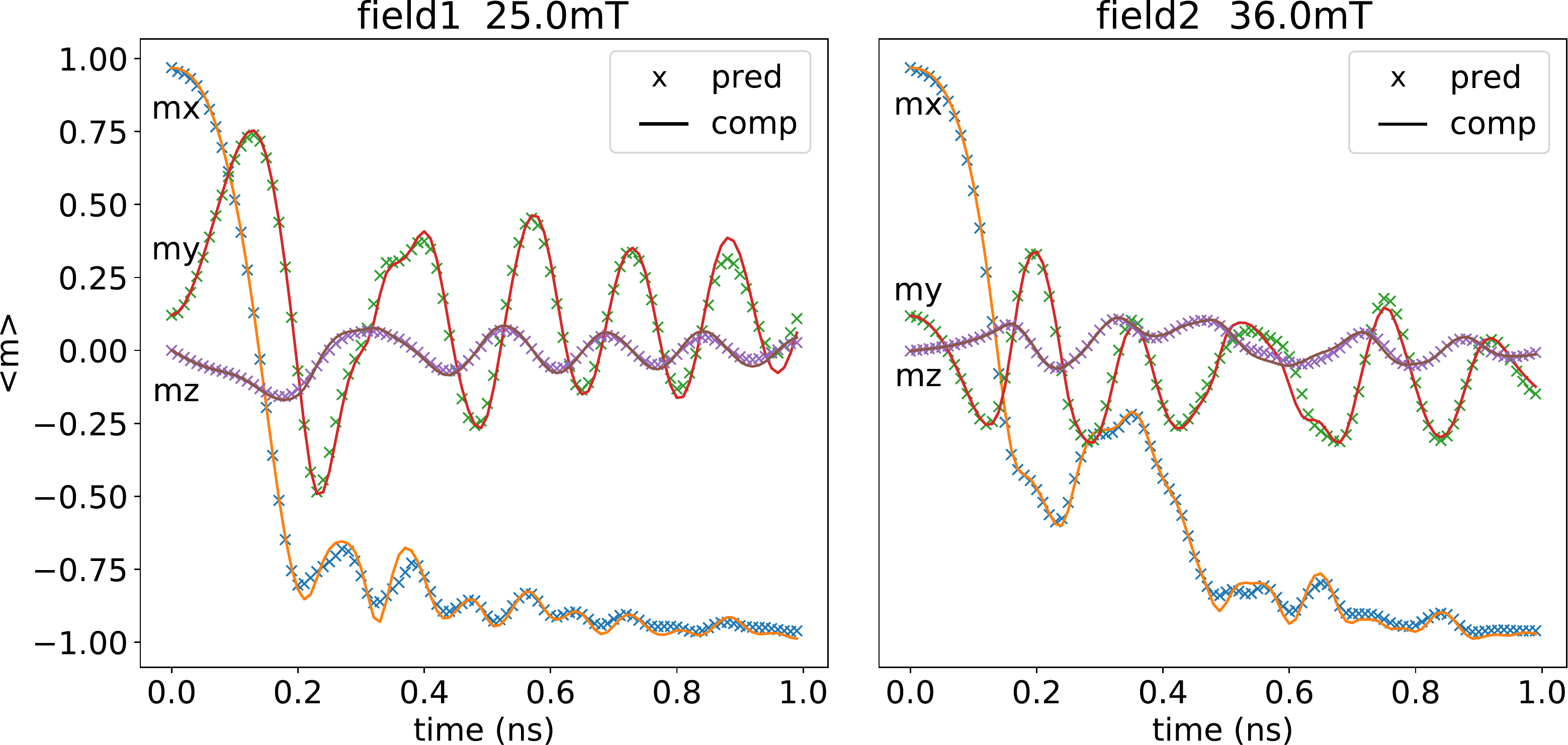}    
\caption{Computed versus predicted mean magnetization components as function of time (ns) for (left) field1: $\mu_0 H_x = 25\cos(170 \pi/180)$mT, $\mu_0 H_y = 25\sin(170 \pi/180)$mT and $H_z = 0$mT and (right) 
field2: $\mu_0 H_x = 26\cos(190 \pi/180)$mT, $\mu_0 H_y = 36\sin(190 \pi/180)$mT and $H_z = 0$mT. Solid lines represent computed values, while the crosses indicate the predicted values. Two kernel principal components were used 
for all time step predictions in each of the two field cases. The system only depends on the magnitude of the field $h$ which was sampled in the range $20-30$mT (field1) and $30-40$mT (field2) for training purpose.}\label{fig:f1a2}
\end{figure}

\begin{figure}[hbtp]
\centering 
\includegraphics[scale=0.14]{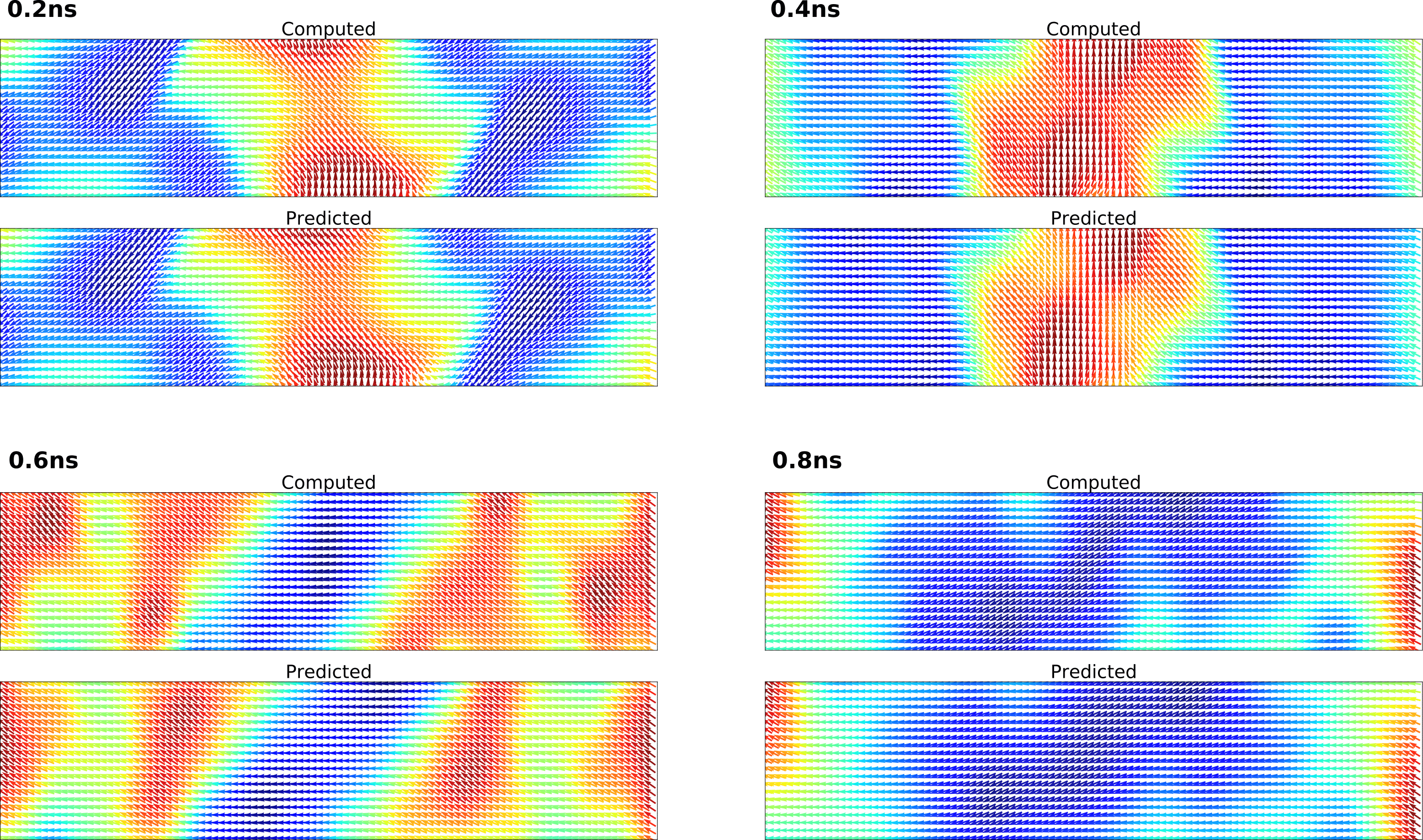}    
\caption{Snapshots of magnetization states at different times for field1 ($\mu_0 H_x = 25\cos(170 \pi/180)$mT, $\mu_0 H_y = 25\sin(170 \pi/180)$mT and $H_z = 0$mT). 
At each time the top images show original states and the bottom approximate pre-images of predicted states in feature space where $2$ principal components were used.}\label{fig:snaps}
\end{figure}

\begin{figure}[hbtp]
\centering 
\includegraphics[scale=0.45]{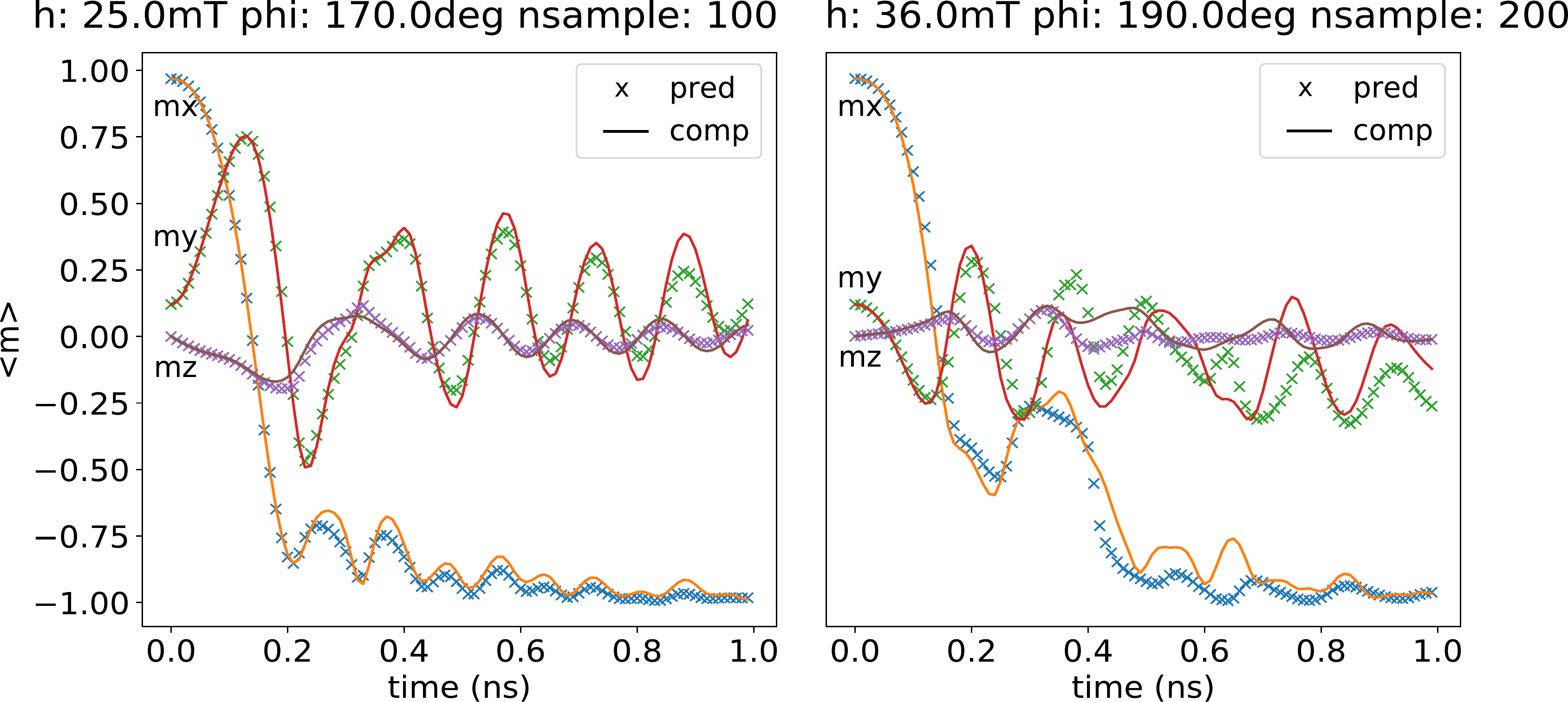}    
\caption{Computed versus predicted mean magnetization components as function of time (ns) for (left) field1: $\mu_0 H_x = 25\cos(170 \pi/180)$mT, $\mu_0 H_y = 25\sin(170 \pi/180)$mT and $H_z = 0$mT and (right) 
field2: $\mu_0 H_x = 26\cos(190 \pi/180)$mT, $\mu_0 H_y = 36\sin(190 \pi/180)$mT and $H_z = 0$mT. Solid lines represent computed values, while the crosses indicate the predicted values. Two kernel principal components were used 
for all time step predictions in each of the two field cases. The system depends on the magnitude of the field $h$ and the in-plane angle, which were randomly sampled in the range $20-30$mT and $160^\circ-180^\circ$ (field1)
and $30-40$mT and $180^\circ-200^\circ$ (field2), respectively, for training purpose.}\label{fig:f1a2_2}
\end{figure}

We look at the NIST $\mu$MAG Standard problem $\#4$ \cite{mumag4}. The geometry is a magnetic thin film of size $500 \times 125 \times 3$ nm with material parameters of permalloy: 
$A = 1.3 \times 10^{-11}$ J/m, $M_s = 8.0 \times 10^5$ A/m, $\alpha = 0.02$. The initial state is an equilibrium s-state, obtained after applying and slowly reducing a saturating field along the diagonal direction $[1,1,1]$ to zero. 
Then two scenarios of different external fields are studied: field1 of magnitude $25$mT is applied with an angle of $170^\circ$ c.c.w. from the positive $x$ axis, field2 of magnitude $36$mT is applied with an angle of $190^\circ$ c.c.w. from the positive $x$ axis. 
For data generation we use a spatial discretization of $100 \times 25 \times 1$ and apply finite differences \cite{miltat2007numerical} to obtain a system of ODEs that is then solved with 
a projected Runge-Kutta method of second order with constant step size of $40$fs.\\ 
\textbf{(i) Dependence on magnitude of field only.} \\
We fix the in-plane angle $\varphi$ of the field to either $170^\circ$ or $190^\circ$. It is known that the benchmark problem depends on the parameters increasingly more discontinuously as larger field values and angles towards the respective values of field2 are attained \cite{mcmichael2001switching}. 
Thus, we will split our data into one part around field1 and one around field2. In the case of field1 we take $100$ values for h equidistantly distributed in $20-30$mT, and analogously for field2 in the range of 
$30-40$mT. We store for each $h$ value in the respective ranges the computed magnetization states every $\Delta t= 0.01$ns, hence in total the complete data set consists of $10^4$ states. 
We include the respective field value $h$ in units of Tesla and scaled with factor $10$ to the states in the data set. 
The maps $f_{t,t + \Delta t}^{(p)}$ are trained using a subset of the complete data set which excludes $5\%$ of the states for validation purpose only on a (uniformly distributed) random basis plus the states associated with the specific field values 
corresponding to field1 and field2, respectively. We use the python implementation of scikit-learn v0.20.3 \cite{scikitlearn} for the kPCA with RBF kernels and the default parameters. 
Finally, the number of principal components were chosen to be $2$ in all tests, while we remark that the choice of only $1$ component yields roughly larger deviations in the experiments below.\\
Fig.~\ref{fig:f1a2} shows computed versus predicted mean magnetization as function of time for both fields. The deviations are acceptably small and compared to \cite{kovacs2019learning} seem to be roughly 
smaller especially in the case of field2. However we here have fixed the angle of the field and thus only have one parameter dependence in the dynamical system.
Fig.~\ref{fig:snaps} shows the computed versus the predicted (approximate pre-images) of magnetization states for different times. 
We observe slight loss of local details for the snapshots corresponding to time $t=0.6$ns and also $t=0.8$ns similar as in \cite{kovacs2019learning}.\\
\textbf{(ii) Dependence on magnitude and in-plane angle.}\\
Now we also vary the in-plane angle and also include the respective value in units of rad  in the data set. We found that scaling the field magnitude with a factor of $100$ and the angle with $1000$ in 
the case of field1, respectively, factors of $10$ for field magnitude and $100$ for angle in the case of field2, gave good results. 
We split the range of training data. For field1 we now take $100$ uniformly random-sampled values $h \in [20,30]$ and $\varphi \in[160^\circ,180^\circ]$ and for field2 we take $200$ sampled as  
$h \in [30,40]$ and $\varphi \in[180^\circ,200^\circ]$. Fig.~\ref{fig:f1a2_2} shows computed versus predicted mean magnetization as function of time for both fields. While the quality of the predictions in the 
field1-case is still good, the predictor in the case of field2 loses accuracy from roughly $0.3$ns onwards. However, the solution of the PDE \eqref{eqn:LLGM} lacks continuous dependency on the external field parameters in this regime, which 
is tough for any prediction method. 
 \begin{figure}[hbtp]
\centering 
\includegraphics[scale=0.8]{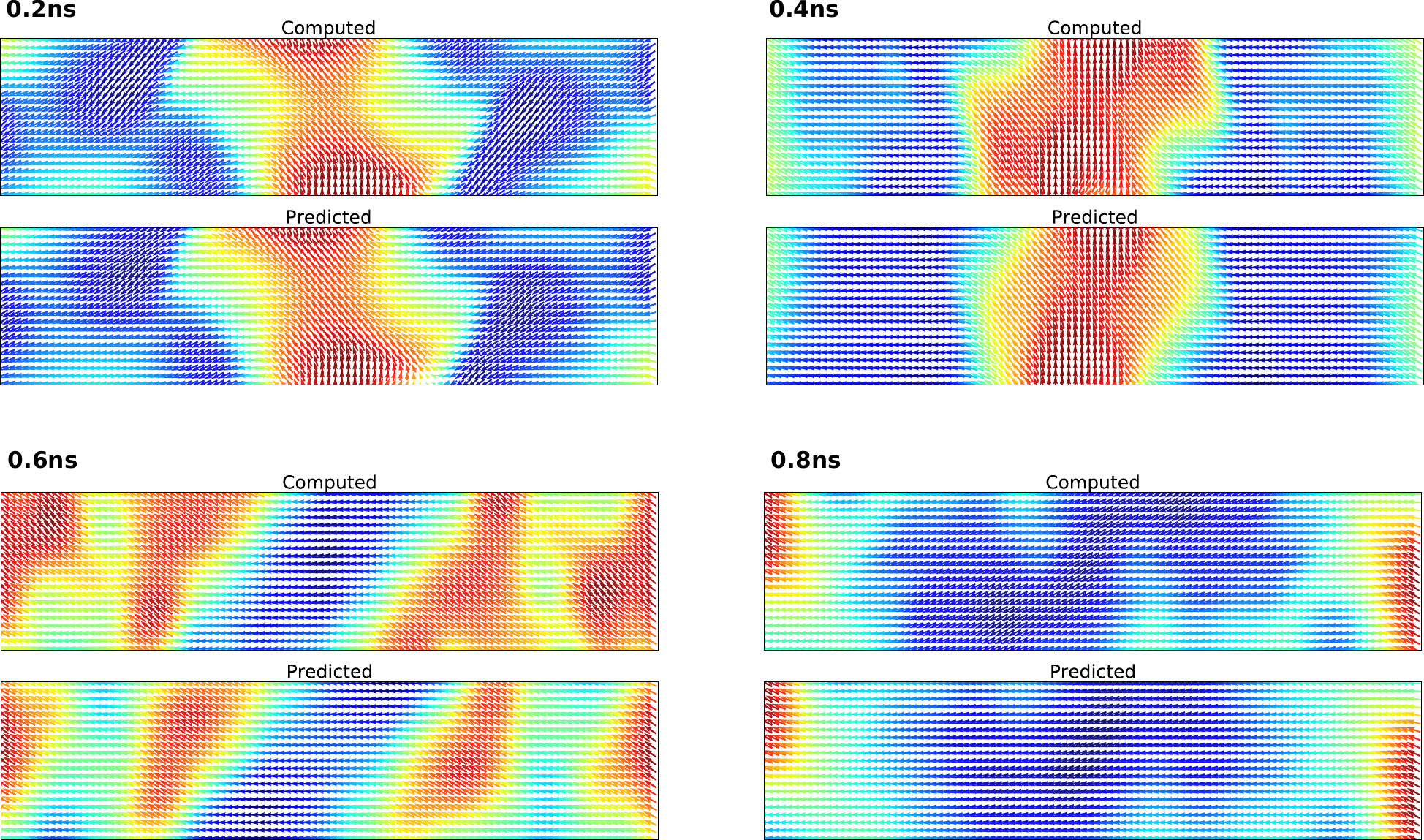}    
\caption{\luk{}{Snapshots of magnetization states in the dynamical system case depending on magnitude and in-plane angle of field at different times for field1 ($\mu_0 H_x = 25\cos(170 \pi/180)$mT, $\mu_0 H_y = 25\sin(170 \pi/180)$mT and $H_z = 0$mT). 
At each time the top images show original states and the bottom approximate pre-images of predicted states in feature space where $2$ principal components were used.}}\label{fig:snaps2}
\end{figure}
\luk{}{Fig.~\ref{fig:snaps2} gives the snapshots of this test example analogue to Fig.~\ref{fig:snaps}.}
\section*{Discussion}
In real applications one would very likely not know the critical behavior of the system and thus could not split the parameter set beforehand. 
Scaling the parameters corresponds to changing the weighting in the components of the arguments of the RBF kernel, which increases the bias in the training set towards the over-weighted components. 
\luk{}{Since the kernel determines the nonlinear map and embedding of the data in the feature space, one can expect changes in the quality of the resulting machine predictor if the kernel is adapted. We found that different scenarios obviously require such an adaption of the kernel to maintain the quality of the reproduced magnetic domain structure. We adjust the scaling ''by hand''.} 

We observed that such scaling of the parameters is critical for the benchmark, whereas in the case of the tougher field2 the quality of the predictions were more sensitive to changes in the scaling. 
A step towards automatization of the training process could be accomplished by tuning parameters of the underlying method e.g. by cross-validation \luk{}{like the authors did in the models for machine learning analysis for microstructures, see \cite{exl2018magnetic} and references therein.}  
However, we emphasize that such \textit{hyper-parameter tuning}, such as for $\gamma$ in the RBF and the scaling factors for the field magnitude and in-plane angle 
are not within the scope of this presentation and part of future investigation. \luk{}{Furthermore, we mention that a data-dependent definition of the kernel function, which is subject of current research, would circumvent the scaling issue because the kernel is adapted to the concrete scenario by definition. }\\

Finally let us remark that the actual computations of the predicted magnetization curves only takes a few seconds of cpu time since the methodology of this data-driven approach 
is to shift computational effort to the precomputation of the training and validation data. In this sense, the proposed method is useful for fast response curve estimation. 

\section*{Conclusion} 
We introduced a time-stepping learning algorithm and applied it to the equation of motion in micromagnetism as a dynamical system depending on the external field as parameter. The approach is based on nonlinear model reduction by means of kernel principal component analysis, a well-known and powerful 
methodology in unsupervised learning. Magnetization states from simulated micromagnetic dynamics associated with different external fields are used to learn a low-dimensional representation in so-called feature space 
and a time-stepping map between the reduced spaces. 
Remarkably, only two principal components in feature space were enough to predict the nonlinear dynamics of the benchmark \cite{mumag4} in both field cases sufficiently well.
Almost all computational effort is shifted to precompute the training and validation data, the training itself and the prediction in the experiments only takes a few seconds.
The approach comes with no restrictions on the spatial discretization such as uniform (finite difference) discretization and might be useful for simulations were computation time is crucial such as 
for sensor applications. Future work could concern the adaption of the proposed time-stepping learning scheme to the prediction of e.g. iterates in energy minimization for permanent magnets applications.   
Finally we remark that the proposed approach of learning time-stepping maps via kernels is not exclusively designed for the LLG equation and thus could be applicable and 
useful also for other parameter-dependent dynamical systems.

\section*{Acknowledgments}
We acknowledge financial support by the Austrian Science Foundation (FWF) via the projects "ROAM" under
grant No. P31140-N32, SFB "ViCoM" under grant No. F41 and SFB "Complexity in PDEs" under grant No.
F65. We acknowledge the support from the Christian Doppler Laboratory Advanced Magnetic Sensing and
Materials (financed by the Austrian Federal Ministry of Economy, 190 Family and Youth, the National
Foundation for Research, Technology and Development). The authors acknowledge the University of Vienna
research platform MMM Mathematics - Magnetism - Materials. The computations were partly achieved by
using the Vienna Scientific Cluster (VSC) via the funded project No. 71140. We thank Georg Kresse for
critically reading the manuscript and a discussion on kPCA applications in materials sciences.
\bibliographystyle{abbrv}
\bibliography{bibref}

\end{document}